\documentclass[12pt,centertags,oneside]{amsart}
\usepackage{amsmath,amstext,amsthm,amscd,typearea,hyperref}
\usepackage{amssymb}
\usepackage{a4wide}
\usepackage[mathscr]{eucal}
\usepackage{mathrsfs}
\usepackage{typearea}
\usepackage{charter}
\usepackage{pdfsync}
\usepackage[a4paper,width=16.2cm,top=3cm,bottom=3cm]{geometry}

\numberwithin{equation}{section}

\newtheorem{theorem}{Theorem}[section]

\newtheorem{proposition}[theorem]{Proposition}
\newtheorem{corollary}[theorem]{Corollary}
\newtheorem{lemma}[theorem]{Lemma}

\newcommand{\cali}[1]{\mathscr{#1}}

\newcommand{\Leb}{{\rm Leb}}

\newcommand{\MZ}{{\rm MZ}}

\renewcommand{\Im}{\mathop{\mathrm{Im}}}
\renewcommand{\Re}{\mathop{\mathrm{Re}}}

\newcommand{\dist}{\mathop{\mathrm{dist}}\nolimits}

\newcommand{\ddc}{dd^c}
\newcommand{\dc}{d^c}

\newcommand{\dbar}{\overline\partial}
\newcommand{\ddbar}{\partial\overline\partial}

\newcommand{\Cc}{\cali{C}}

\newcommand{\B}{\mathbb{B}}
\newcommand{\D}{\mathbb{D}}
\newcommand{\C}{\mathbb{C}}

\newcommand{\N}{\mathbb{N}}

\newcommand{\R}{\mathbb{R}}

\newcommand{\fM}{\mathfrak{M}}
\newcommand{\fR}{\mathfrak{R}}
\newcommand{\fZ}{\mathfrak{Z}}

\title[Distribution of scattering resonances]{Distribution of scattering resonances for generic\break Schr\"odinger operators}

\author{Tien-Cuong Dinh}
\address{Department of Mathematics, National University 
of Singapore, 10 Lower Kent Ridge Road, Singapore 119076. {\tt http://www.math.nus.edu.sg/$\sim$matdtc}}
\email{matdtc@nus.edu.sg}
 
\author{Vi{\^e}t-Anh Nguy{\^e}n}
\address{Universit\'e de Lille 1, 
Laboratoire de math\'ematiques Paul Painlev\'e, 
CNRS U.M.R. 8524,  
59655 Villeneuve d'Ascq Cedex, 
France. {\tt http://www.math.univ-lille1.fr/$\sim$vnguyen}}
\email{Viet-Anh.Nguyen@math.univ-lille1.fr}
\medskip

\date{September 17, 2017}

\begin{document}

\maketitle

\begin{abstract}
Let $-\Delta+V$ be  the Schr\"odinger operator acting on $L^2(\R^d,\C)$ with $d\geq 3$ odd. Here $V$ is a bounded real- or complex-valued function vanishing outside the
closed ball of center $0$ and radius $a$. If $V$ belongs to the  class $\fM_a$ of potentials introduced by Christiansen, we show that 
when $r\to\infty$, the resonances of  $-\Delta+V,$  scaled down by the factor $r$, are asymptotically distributed, with respect to an explicit probability distribution on 
the closed lower unit half-disc of the complex plane. The rate of convergence is also considered for subclasses  of potentials.
\end{abstract}

\noindent
{\bf Classification AMS 2010}: 35P25, 47A40. 

\noindent
{\bf Keywords: } Schr\"odinger operator, resonance, scattering matrix, scattering pole.

\section{Introduction} \label{S:intro}

Let $\Delta$ denote the Laplacian operator on $\R^d.$ In this  work, we only consider  $d$ odd because  the  case with $d$ even is  of another nature. Let $V$ be a bounded complex-valued function with support in the closed ball $\overline \B_a$ of  center $0$ and 
radius $a$ in $\R^d$, that is, $V\in L^\infty(\overline\B_a,\C).$
The purpose of this work is to establish   the   distribution law  for the resonances associated to the Schr\"odinger operator $-\Delta+V$ acting on $L^2(\R^d,\C)$
for  ``most of"   potentials in $L^\infty(\overline\B_a,\C)$,
or in $L^\infty(\overline\B_a,\R)$ if  we only consider  real  potentials.
The study of the asymptotic behavior  of resonances  has a long history and was intensively investigated during the last three decades. The reader can find in 
\cite{Christiansen05, Christiansen06b,DZ, Sjostrand, Stefanov,Vodev01, Zworski94, Zworski02} and the references therein an introduction to the subject.

Consider  the complex parameter $\lambda\in \C.$  If $\lambda$ is large enough with $\Im(\lambda)>0$, the operator $R_V(\lambda):=(-\Delta+V-\lambda^2)^{-1}$ on $L^2(\R^d,\C)$ is well-defined and is bounded.
It depends holomorphically on $\lambda$. 
If $\chi$ is any smooth function with compact support such that $\chi V=V$, one can extend $\chi R_V(\lambda)\chi$ to a family of operators which depends meromorphically on $\lambda\in\C$. The poles of this family, which 
are called {\it the resonances} of the operator $-\Delta+V$,
and their multiplicities do not depend on the choice of $\chi$.  
Let $\fR_V$ denote the set of resonances of $-\Delta+V$, where each element is counted according to its multiplicity. 
Denote by $n_V(r)$ the number of resonances of modulus $\leq r$ counted with multiplicity.

In dimension $d=1$, Zworski obtained in \cite{Zworski87} the following   estimate 
\begin{equation*} 
n_V(r)={4\over \pi} ar  +o(r) \quad \mbox{as}\quad r\to\infty,
\end{equation*}
where $2a$ is the diameter of the support of $V$, see also \cite{Froese, Regge, Simon,Zworski02}. In this  paper we only consider the  dimension  $d\geq 3$.

The upper bound for the number of resonances is well-understood
while, in contrast, the lower bound  is still not completely understood.
Set
\begin{equation}\label{e:N}
N_V(r):=\int_0^{r} {n_V(t)-n_{V}(0)\over t} dt.
\end{equation}
We have
\begin{equation}\label{e:Melrose-Zworski}
dN_V(r)\leq c_da^dr^d+O(r^{d-1}\log r) \quad \mbox{as}\quad r\to\infty,  
\end{equation}
where  $c_d$ is defined in Section \ref{s:measures}. This constant $c_d$ is sharp and  
was identified by Stefanov in \cite{Stefanov}.
The last estimate is a direct consequence of Propositions \ref{P:Dinh-Vu} and \ref{P:Christiansen} below. It generalizes an estimate of 
Zworski in \cite{Zworski89b} where he obtained $o(r^d)$ instead of   $O(r^{d-1}\log r)$. See also \cite{SZ, Vodev92} for more general results and \cite{Melrose82, Melrose83, Melrose84} for earlier results.

Let $0<\nu \leq 1$ be a constant. The following family of potentials was  introduced   in  \cite{DinhVu}
$$\fM^\nu_a:=\Big\{V\in L^\infty(\overline\B_a,\C):\  n_V(r)-c_da^dr^d=O(r^{d-\nu+\eta}) \mbox{ as } r\to\infty \mbox{ for every } \eta>0\Big\}.$$
Clearly, this is a subset of the following family introduced earlier by Christiansen in \cite{Christiansen12}
$$\fM_a:=\Big\{V\in L^\infty(\overline\B_a,\C): \   n_V(r)-c_da^dr^d=o(r^d) \mbox{ as } r\to\infty\Big\}.$$
We will call $\fM_a$  {\it Christiansen class}. By the  results of \cite{Stefanov,Zworski89a}, $\fM_a$ contains all radial real-valued functions $V(z)=V(\|z\|)$ of class $\Cc^2$ on $\overline\B_a$
with $V(a)\not=0.$ 
In \cite{Christiansen06a} Christiansen exhibits an  example of a  smooth  complex-valued potential on $\overline\B_a$ which  does not vanish  on $b\B_a$ such that  $\fR_V$ is  empty. 
Such   a  function does not belong   to $\fM_a.$
Moreover,  Vu and the first   author    proved in \cite{DinhVu} that
generic potentials in $L^\infty(\overline\B_a,\C)$ or    in   $L^\infty(\overline\B_a,\R)$ belong to $\fM^{3/16}_a$, see also \cite{Christiansen99, CH1, Sa, SaZ}.

Consider a potential $V$ in $\fM_a$ and define for $r>0$
\begin{equation}\label{e:mu_R_V,r}
\mu^\fR_{V,r}:={1\over c_d a^d r^d }\sum_{z\in\fR_V} \delta_{z/r}, 
\end{equation}
where $\delta_{z/r}$ denotes the Dirac mass at the point $z/r$.
The fact that $V$ belongs to $\fM_a$ implies that the mass of $\mu^\fR_{V,r}$ on the unit disc tends to 1 as $r$ tends to infinity.

We will introduce later in Section \ref{s:measures} the positive constants $c_d,e_d$ and the function $h_d:[0,\pi]\to\R_+$.
Consider the measure $\mu^0$ with support in $\R$ which is absolutely continuous and has density  ${e_d\over 2\pi  c_d}|x|^{d-1}$ with respect to the Lebesgue measure on $\R$. 
Consider also the measure $\mu^-$ with support in the lower half-plane $\C_-$ which is absolutely continuous with respect to the Lebesgue measure on $\C$ and has density
$$\kappa(z):={1\over 2\pi c_d} r^{d-2} \big[d^2h_d(|\theta|) +h''_d(|\theta|)\big] \qquad \text{with} \quad z=re^{i\theta},\ r=|z|,\ \theta\in (-\pi,0).$$
We define $\mu_\MZ:=\mu^0+\mu^-$ and call it {\it Melrose-Zworski distribution}. It will be shown later that $\mu_\MZ$ is a positive measure vanishing on the open upper half-plane $\C_+$. Moreover,  its restriction to the unit disc is a probability measure. 
The Melrose-Zworski distribution is homogeneous of degree $d$ : 
if $A_\lambda:\ \C\to\C$ denotes  the  dilation $z\mapsto \lambda z$ with $\lambda>0$, 
then  $(A_\lambda)^*(\mu_\MZ)= \lambda^d\mu_\MZ.$

For   a set $W\subset \C$ and  a  number $r>0,$
let $rW$ denote  the dilation of $W$ by $r,$ that is,
$rW:=\left\lbrace rz:\ z\in \Omega\right\rbrace$. 
Let  $n_{V,W}(r)$ be the  number  of resonances,  counted   with multiplicity, in $rW.$   
In particular, for $W=\D$  we have   $n_{V,\D}(r)=n_V(r).$
Our first main result is the following theorem.
 
 \begin{theorem} \label{T:main_1}
 Let $V$ be a function in the Christiansen class $\fM_a$. Then, the Schr\"odinger operator $-\Delta+V$ satisfies the Melrose-Zworski resonance law : when $r\to\infty$,
 the above-defined measure $\mu^\fR_{V,r}$ converges weakly to the Melrose-Zworski distribution $\mu_\MZ$  
    on $\C$, i.e.
 $$\int \varphi  d  \mu^\fR_{V,r}\to  \int\varphi  d  \mu_\MZ\quad\text{as}\quad  r\to\infty,$$
 for all  continuous   functions  $\varphi$  with   compact support in $\C.$
 In particular,  for every bounded open set $\Omega$ such that $b\Omega$ has zero area and $b\Omega\cap\R$ has zero length,  we have 
$$n_{V,W}(r)= \mu_\MZ(\Omega) c_d a^d r^d +o(r^d)  \qquad\text{as}\qquad r\to\infty$$
for any set $W$ such that $\Omega\cap\C_-\subset W\subset\overline \Omega$, e.g. $W=\Omega\cap\C_-, \Omega\cap\overline\C_-,\Omega$ or $\overline\Omega$. 
 \end{theorem}
 
 Consider now the following  family  of open  sectors  $\Omega(\theta_1,\theta_2)$ in the lower half-plane indexed  by $0\leq\theta_1<\theta_2\leq \pi:$  
 $$
 \Omega(\theta_1,\theta_2):=\big\lbrace  z \in \D:\  \theta_1-\pi < \arg z < \theta_2-\pi  \big\rbrace.
 $$
We will  applying Theorem  \ref{T:main_1}   to these windows and obtain the following   result, see \cite[Prop. 1.1  and  Cor. 1.4]{Christiansen12}, see also 
\cite{SZ2,Stefanov0,Zerzeri} for related results.
 
 \begin{corollary}[Christiansen]\label{C:Christiansen}
 For $\Omega:=\Omega(\theta_1,\theta_2),$
    we have 
 \begin{equation*}
  n_{V,\Omega}(r)= c(\theta_1,\theta_2) c_d a^d r^d +o(r^d)\quad \text{and} \quad n_{V,\overline\Omega}(r)= c(\theta_1,\theta_2) c_d a^d r^d +o(r^d)\quad \text{as}\quad r\to\infty,
  \end{equation*}
 where
 $$ c(\theta_1,\theta_2):={1\over 2\pi dc_d}\Big[c(\theta_2)-c(\theta_1)  +d^2\int_{\theta_1}^{\theta_2} h_d(\theta)d\theta  \Big]  
 \quad \text{and} \quad
 c(\theta):= \begin{cases}
 h_d'(\theta)  &  \text{for}\ 0 <\theta< \pi \\
0   &  \text{for}\ \theta=0\ \text{or}\ \pi.
  \end{cases}
$$
 \end{corollary}
 
 We can      prove   Theorem \ref{T:main_1}  using the last  result  and  some standard  techniques.
 However,  we  will consider  in this paper  a  more direct and   simpler  approach.
 The novelties of  our  approach is  that   it not only  gives us   an explicit measure (Melrose-Zworski distribution), but also leads us  to   effective   estimates of the  rate of the convergence  which  will be   presented in the next result.

 Let $\Omega$   be  any open  set in $\C$  and  let $\gamma > 0$ be a positive number.    Given   $\mu$ and $\mu'$  two positive measures  on $\Omega,$  define 
\begin{equation} \label{e:dist-gamma}
\dist_{\Omega,\gamma} (\mu, \mu' ) := \sup  |\langle \mu -\mu' ,\varphi \rangle|,
\end{equation}
where the pairing  $\langle \mu -\mu' , \varphi\rangle$ denotes the integral of $\varphi$ with respect to the measure $\mu-\mu'$
and the supremum is taken over all $\Cc^\gamma$ functions $\varphi$ with compact support in $\Omega$ with
$\|\varphi\|_{\Cc^\gamma} \leq 1$.
As in \cite{DinhSibony10}, if $\gamma,\gamma'$ and $\Omega,\Omega'$ satisfy $0 <\gamma \leq\gamma' $ and $\Omega\Subset \Omega'\subset \C$, on any subset of measures whose masses on $\Omega'$ are bounded by a constant, we have the following inequalities for  some constant $c > 0$ 
\begin{equation} \label{e:dist-gamma-gamma}
\dist_{\Omega,\gamma'} \leq \dist_{\Omega, \gamma} \leq c[\dist_{\Omega',\gamma'} ]^{\gamma/\gamma'} .
\end{equation}
The function $\dist_{\Omega, \gamma}$ is  a semi-distance on the space of  locally  finite positive measures on $\C.$ Note that $\dist_{\Omega, 1}$ is related to the well-known
Kantorovich-Wasserstein distance for measures. 

\begin{theorem} \label{T:main_2}
 Let $V$ be a function in the  class $\fM^\nu_a$  for some $0<\nu\leq 1,$   let $\eta>0$ be an arbitrary constant and let $\Omega$ be a  bounded domain in $\C.$ 
Then
  for every number $0<\gamma\leq 1,$  we have that
 $$\dist_{\Omega,\gamma}(\mu^\fR_{V,r},\mu_\MZ)\leq  c r^{-\gamma\nu/2+\eta} \quad \text{for } r \text{ large enough},$$
 where   $c>0$ is a constant  which depends   on $a,V, \gamma,\nu$ and $\eta$ but is independent of $r$.
Moreover, if  the boundary of $\Omega$ is  piecewise smooth   and   transverse to the real line $\R$, then 
$$ n_{V,W}(r)= \mu_\MZ(\Omega)  c_da^d r^d  +o(r^{d-\nu/3+\eta})\quad\text{as}\quad r\to\infty$$
for any set $W$ such that $\Omega\cap\C_-\subset W\subset\overline \Omega$, e.g. $W=\Omega\cap\C_-, \Omega\cap\overline\C_-,\Omega$ or $\overline\Omega$. 
 \end{theorem}

\medskip
\noindent
{\it Notation and convention.} 
Denote by $\B_a$ the open ball of center $0$ and  radius $a$ in $\R^d.$
For   a set $\Omega\subset \C$ and $r>0,$ let $r\Omega:=\left\lbrace rz:\ z\in \Omega\right\rbrace$  and let $b\Omega$  denote the boundary of $\Omega.$
Let $\D$  (resp. $\D(s)$) be the open unit disc (resp. the disc of center $0$ and radius $s$) in $\C.$  
Define $\C_\pm := \{z \in \C :\ \pm \Im z > 0\}$ and  $\D_\pm:= \D\cap \C_\pm.$ The function $h_d$ and the constant $c_d $  are introduced in Section \ref{s:measures}.
Write $\arg z := \theta$ and $\log z := \log r + i\theta$ for $z = re^{i\theta}$ with $r > 0$ and $\theta \in [-\pi, \pi ].$
Recall that $\dc:={1\over 2\pi i}(\partial-\dbar)$ and  $\ddc={i\over \pi}\ddbar$.
Both $\Leb$ and $dxdy$ (resp. $\Leb_\R$ and  $dx$) denote the Lebesgue measure on $\C$ (resp. on $\R$), where we use the canonical coordinates $z=x+iy$.
Let $L^\infty(\overline\B_a,\C)$ (resp.  $L^\infty(\overline\B_a,\R)$)  be the  space of all  bounded complex- (resp. real-) valued functions  with    support in $\overline\B_a$.
The constants we use can be changed from line
to line but they are independent of $r$. The notations $\lesssim$ and $\gtrsim$ mean inequalities up to a multiplicative constant. 

\medskip
\noindent
{\bf Acknowledgement.} The  first author  was supported by Start-Up  Grant R-146-000-204-133 
from the National University of Singapore. The paper was partially prepared 
during the visit  of the first author at the Freie Universit\"at Berlin
and of the second author at the National University of Singapore. They would like to thank these organizations, the Alexander von Humboldt foundation and H\'el\`ene Esnault for their hospitality and  support.

\section{Properties of some positive measures}\label{s:measures}

In this section, we will give basic results on positive measures in $\C$ and their potentials, see \cite{DNS, LelongGruman, Ransford}. The measures we consider are locally finite Borel measures. We also study some properties of
the Melrose-Zworski distribution that will be used later in the proof of our main results.

  \begin{lemma}\label{l:simple_cv}
      Let $\mu_k$, with $k\in\N$, and $\mu$ be  positive  measures in $\C$ such that  $\mu_k$ converges to $\mu$ weakly as $k$ tends to $\infty$.
   Let $\Omega$ be  a bounded opens set in $\C$ and assume  that $\mu(b\Omega)=0$. 
   Then  $\mu_k(W)\to\mu(\Omega)$ as $k\to\infty$ for every set $W$ such that $\Omega\subset W \subset \overline \Omega$.
   \end{lemma}
\proof
It is enough to prove that $\mu_k(\Omega)\to\mu(\overline \Omega)$  and $\mu_k(\overline\Omega)\to\mu(\Omega)$.
Choose two sequences of continuous functions $0\leq \chi_n\leq \rho_n\leq 1$ with compact support in $\C$ such that $\chi_n$ increases to the characteristic function of $\Omega$ and $\rho_n$ decreases to the characteristic function of $\overline\Omega$. We have for each $n$ 
$$\liminf_{k\to\infty} \mu_k(\Omega) \geq \liminf_{k\to\infty} \langle \mu_k,\chi_n\rangle = \langle \mu,\chi_n\rangle.$$
Taking $n\to\infty$ gives 
$$\liminf_{k\to\infty} \mu_k(\Omega)\geq \mu(\Omega) =\mu(\overline\Omega).$$
We use here that $\mu(b\Omega)=0$.
Similarly, we have
$$\limsup_{k\to\infty} \mu_k(\overline\Omega) \leq \limsup_{k\to\infty} \langle \mu_k,\rho_n\rangle = \langle \mu,\rho_n\rangle.$$
Taking $n\to\infty$ gives 
$$\limsup_{k\to\infty} \mu_k(\overline\Omega)\leq \mu(\overline \Omega) =\mu(\Omega).$$
Therefore, we get $\mu_k(\Omega)\to\mu(\Omega)$ and $\mu_k(\overline\Omega)\to\mu(\Omega)$. This implies the lemma.
 \endproof
 
We will give now some results which allow us to get the rate of convergence of positive measures on $\C$. Let  $\Omega$ be an open set in $\C$, not necessarily bounded, such that $b\Omega$ is compact. 
For $\epsilon>0$, denote by $(b\Omega)_\epsilon$ the set of points $z$ of distance less than $\epsilon$ to $b\Omega$. Let $\vartheta_\Omega(\epsilon)$ be the infimum of the numbers $\vartheta>0$ such that 
\begin{enumerate}
\item[-] there is a function $0\leq \chi\leq 1$ of class $\Cc^2$ with support in $\Omega$ and equal to 1 on $\Omega\setminus {(b\Omega)}_\epsilon$ such that $\|\chi\|_{\Cc^2}\leq\vartheta$; 
\item[-] there is a function $0\leq \rho\leq 1$ of class $\Cc^2$ with support in $\Omega\cup (b\Omega)_\epsilon$ and equal to 1 on $\overline\Omega$ such that $\|\rho\|_{\Cc^2}\leq\vartheta$.
\end{enumerate} 
If such functions do not exist, we define $\vartheta_\Omega(\epsilon):=+\infty$.

\begin{lemma} \label{l:dist}
Let $u_1$ and $u_2$ be two subharmonic functions on an open set $U$ in $\C$. Define $\mu_1:=\ddc u_1$ and $\mu_2:=\ddc u_2$. Let $\Omega$ be an open set in $\C$ such that $\Omega\Subset U$ and let $0<\gamma\leq 2$ be a constant. Then there is a constant $c>0$  depending only on $\Omega,U,\gamma$ such that 
$$\dist_{\Omega,\gamma}(\mu_1,\mu_2)\leq c \|u_1-u_2\|_{L^1(U)}^{\gamma/2} \quad \text{and} \quad |\mu_1(\Omega)-\mu_2(\Omega)| \leq \mu_1((b\Omega)_\epsilon) + c\vartheta_\Omega(\epsilon)\|u_1-u_2\|_{L^1(U)}$$
for $\epsilon>0$ small enough. Moreover, the last estimate still holds if we replace $\mu_1(\Omega)-\mu_2(\Omega)$ by $\mu_1(\overline\Omega)-\mu_2(\overline\Omega)$.
\end{lemma}
\proof
We first prove the first inequality for $\gamma=2$. Let $\varphi$ be any $\Cc^2$ function with compact support in $\Omega$ such that $\|\varphi\|_{\Cc^2}\leq 1$. By Stokes formula, we have
$$|\langle \mu_1-\mu_2,\varphi\rangle| = |\langle \ddc (u_1-u_2),\varphi\rangle| = |\langle u_1-u_2,\ddc\varphi\rangle| = \Big|\int_\Omega (u_1-u_2)\ddc\varphi\Big|.$$  
Since $\ddc\varphi$ is a bounded differential form, the last integral is bounded by a constant times $\|u_1-u_2\|_{L^1(U)}$. By \eqref{e:dist-gamma}, the first estimate in the lemma holds for $\gamma=2$.
If $\Omega'$ is an open set such that $\Omega\Subset\Omega'\Subset U$, we obtain in the same way that $\dist_{\Omega',2}(\mu_1,\mu_2)$ is bounded by a constant times $\|u_1-u_2\|_{L^1(U)}$. This, together with \eqref{e:dist-gamma-gamma}, imply the first assertion in the lemma for every $0<\gamma\leq 2$.

Fix an arbitrary constant $\vartheta>\vartheta_\Omega(\epsilon)$ for $\epsilon>0$ small enough. Let $\chi$ and $\rho$ be as above. We have
$$\mu_1(\Omega)-\mu_2(\Omega)\leq \mu_1(\Omega)-\langle\mu_2,\chi\rangle \leq \mu_1((b\Omega)_\epsilon) +  \langle\mu_1,\chi\rangle- \langle\mu_2,\chi\rangle.$$
Recall that $\|\chi\|_{\Cc^2}\leq \vartheta$. As it was done for $\varphi$ above, we obtain that $\langle\mu_1,\chi\rangle- \langle\mu_2,\chi\rangle$ is bounded from above by a constant times $\vartheta \|u_1-u_2\|_{L^1(U)}$. Therefore, we have for some constant $c>0$
$$\mu_1(\Omega)-\mu_2(\Omega) \leq \mu_1((b\Omega)_\epsilon) + c\vartheta\|u_1-u_2\|_{L^1(U)}.$$
Similarly, using the function $\rho$, we get 
$$\mu_1(\Omega)-\mu_2(\Omega) \geq -\mu_1((b\Omega)_\epsilon) - c\vartheta\|u_1-u_2\|_{L^1(U)}.$$
This implies the second estimate in the lemma. Note that the same proof holds when we replace $\Omega$ by $\overline\Omega$.
\endproof 
 
Later, we will use the last lemma for $\Omega$ a bounded open set with piecewise smooth boundary. The following result is then useful.
We say that an open set $\Omega$ with compact boundary is {\it nice} if $\vartheta_\Omega(\epsilon)=O(\epsilon^{-2})$ as $\epsilon$ tends to 0. 

\begin{lemma}\label{l:domain} 
Let $\Omega$ be a bounded simply connected open set in $\C$ with piecewise smooth boundary. Assume that the angle at each singular point of its boundary is strictly smaller than $\pi$. Then $\Omega$ is nice. 
\end{lemma}
\proof
Consider first the case of  smooth boundary.
We can find  a   defining smooth function  $\tau:\C\to\R$ 
 such that $\Omega:=\{\tau<0\}$ and that $d\tau\not=0$ on $b\Omega.$ 
 Fix  a  smooth function $\chi_0:\ \R\to [0,1]$  such that
 $\chi_0=1$ on $(-\infty,0]$ and   $\chi_0=0$  on  $[1/2,\infty).$   Fix also two constants $\epsilon_0$ small enough and $A>0$ large enough.
For every $0<\epsilon\leq\epsilon_0$, we can check that  the  functions
 $$\chi(z):= \chi_0   (A\epsilon^{-1}\tau(z)+1) \qquad \text{and} \qquad \rho(z):= \chi_0   (A\epsilon^{-1}\tau(z)-1) $$  
 satisfy the conditions required in the definition of $\vartheta_\Omega(\epsilon)$. Their $\Cc^2$ norms are bounded by a constant times $\epsilon^{-2}$. This implies the result for the smooth case. Note that the $\Cc^1$ norms of $\chi$ and $\rho$ are bounded by a constant times $\epsilon^{-1}$.

Consider the general case. Observe that we can find a finite number of simply connected bounded open sets $\Omega_1,\ldots,\Omega_k$ with smooth boundaries whose intersection is equal to  $\Omega$. Moreover, 

\smallskip
(1) Each smooth piece of $b\Omega$ is contained in $b\Omega_i$ for exactly one index $i$ and conversely, for each $i$,  
$b\Omega_i$ contains exactly one smooth piece of $b\Omega$;

\smallskip
(2) For $i\not=j$, $b\Omega_i$, $b\Omega_j$ intersect  at exactly 2 points and the intersection is transversal. 

\smallskip
(3) For all distinct indexes $i,j,l$, the intersection of $b\Omega_i$, $b\Omega_j$ and $b\Omega_l$ is empty.

\smallskip\noindent
To see this point, we can use a smooth diffeomorphism of $\C$ in order to reduce the problem to the case of a convex polygon.

Fix a constant $c>0$ small enough. We only need to consider $\epsilon>0$ small enough and define $\epsilon':=c\epsilon$. 
For  each $j=1,\ldots, k$, we can
choose the functions $\chi_j$ and $\rho_j$ as in the definition of $\vartheta_{\Omega_j}(\epsilon')$  associated with $\Omega_j, \epsilon'$ instead of $\Omega,\epsilon$ and such that $\|\chi_j\|_{\Cc^2}=O(\epsilon^{-2})$ and $\|\chi_j\|_{\Cc^1}=O(\epsilon^{-1})$. 
Set  $\chi:=\chi_1\ldots\chi_k$ and  $\rho:=\rho_1\ldots\rho_k$.
We can check that these functions satisfy the conditions required in the definition of $\vartheta_\Omega(\epsilon)$ as in the case where $\Omega$ is a convex polygon. 
Moreover, both  $\|\chi\|_{\Cc^2}$ and $\|\rho\|_{\Cc^2}$ are bounded by a constant times
$\epsilon^{-2}$. So $\Omega$ is a nice open set.
 \endproof
 
We will also need the  following  auxiliary  results.
Consider a domain $\Omega$ in $\C$ which is symmetric with respect to the real line $\R$. Define $\Omega_+:=\Omega\cap \C_+$, $\Omega_-:=\Omega\cap \C_-$ and $L:=\Omega\cap\R$.
 
\begin{lemma} \label{l:u0}
Let $u$ be a subharmonic function on $\Omega_+$ such that $u\geq 0$ and $u(z)$ tends to $0$ when $z$ tends to $L$. Then the function 
\begin{equation*}
u_0(z):=
\begin{cases}
 u(z) &  \text{for } z\in\Omega_+\\
 0 & \text{for } z \in \Omega_-\cup L.
\end{cases}
\end{equation*}
is subharmonic on $\Omega$.
\end{lemma}
\proof
Consider for $\epsilon>0$ the function 
\begin{equation*}
u_\epsilon(z):=
\begin{cases}
 \max(u(z),\epsilon) &  \text{for } z\in\Omega_+\\
 \epsilon & \text{for } z \in \Omega_-\cup L.
\end{cases}
\end{equation*}
Clearly, this function is subharmonic in $\Omega_+\cup\Omega_-$ because the maximum of two subharmonic functions is subharmonic. Since subharmonic functions are upper semi-continuous, $u_\epsilon=\epsilon$ in a neighbourhood of $L$. So it is also subharmonic in a neighbourhood of $L$. It follows that $u_\epsilon$ is subharmonic on $\Omega$. When $\epsilon$ decreases to 0, $u_\epsilon$ decreases to $u_0$. Therefore, $u_0$ is also subharmonic. 
\endproof 
 
\begin{lemma}\label{l:subharmonic}
Let $u$ be a continuous function on $\Omega_+\cup L$ such that ${\partial u\over \partial y}$ exists and is continuous there. 
Assume that $u$ is subharmonic on $\Omega^+$.
Define the function  $\tilde u:\ \Omega\to\R$   by 
\begin{equation*}
\tilde u(z):=
\begin{cases}
 u(z) &  \text{for } z\in\Omega_+\cup L\\
 u(\bar{z}) & \text{for } z \in \Omega_-.
\end{cases}
\end{equation*}
\begin{enumerate}
\item[(1)] The function $\tilde u$ is subharmonic in $\Omega$ if and only if ${\partial u\over \partial y}(z)\geq 0$ for all $z\in L$.
\item[(2)]  We have that  $\ddc\tilde u$ is a measure on $\Omega$. Moreover, its
 restriction to $L$ is absolutely continuous 
 and has density ${1\over \pi} {\partial u\over \partial y}(x)$
 with respect to the Lebesgue measure on $L$.
 \end{enumerate}
In particular, we have $\ddc|y|={1\over\pi}\Leb_{\R}$ on $\C$.
\end{lemma}
\proof
(1) Assume that $\tilde u$ is subharmonic. We show that ${\partial u\over \partial y}(z)\geq 0$ for all $z\in L$. Assume
by contradiction that  ${\partial u\over \partial y}(a) <0$ at some point $a\in L$. For simplicity, we can suppose $a=0$.
Let $\rho\geq 0$ be a smooth function on $\R$ with compact support and with integral 1. Consider the following functions obtaining by a convolution with  $\rho^\epsilon(t):=\epsilon^{-1}\rho(\epsilon^{-1}t)$ :
$$u^\epsilon(z):=(\rho^\epsilon \ast u)(z) := \int_\R u(z+t)\rho^\epsilon(t)dt$$
and
$$\tilde u^\epsilon(z):=(\rho^\epsilon \ast \tilde u)(z) := \int_\R \tilde u(z+t)\rho^\epsilon(t)dt.$$
For $\epsilon$ small enough, these functions satisfy similar properties as $u$ and $\tilde u$ do. In particular, $\tilde u^\epsilon$ is subharmonic near 0 and ${\partial u^\epsilon \over \partial y}(0) <0$ because
$${\partial u^\epsilon\over \partial y}(z)=\big(\rho^\epsilon \ast {\partial u\over \partial y}\big)(z) = \int_\R {\partial u\over \partial y}(z+t)\rho^\epsilon(t)dt.$$
Moreover, the restrictions of $u^\epsilon$ and $\tilde u^\epsilon$ to $\R$ are smooth.
So we can replace $u,\tilde u$ by $u^\epsilon, \tilde u^\epsilon$ in order to assume that $u$ is smooth on $L$.

Adding to $\tilde u$ a suitable (harmonic) affine function in $x$ allows us to assume that  $u(0)=0$ and ${\partial u\over \partial x}(0) =0$.
Fix small enough constants $\delta>0$ and  $r>0$ such that ${\partial u\over \partial y}(z) \leq -2\delta$ for $z\in \Omega_+\cup L$ with $|z|\leq r$ and $u(z)\leq \delta |z|$ for $z\in [-r,r]$. For $\theta\in [0,\pi]$, we have using the above property of  ${\partial u\over \partial y}(z)$ that
$$u(re^{i\theta})- u(r\cos\theta) \leq -2\delta |re^{i\theta}-r\cos\theta| =-2\delta r\sin\theta$$
hence
$$u(re^{i\theta}) \leq u(r\cos\theta) -2\delta r\sin\theta\leq \delta r (|\cos\theta| -2\sin\theta).$$
So we have for $\theta\in [-\pi,\pi]$
$$\tilde u(re^{i\theta}) \leq \delta r (|\cos\theta| -2|\sin\theta|).$$
This contradicts the following submean inequality for subharmonic functions
$$\tilde u(0) \leq {1\over 2\pi} \int_{-\pi}^\pi \tilde u(re^{i\theta}) d\theta$$
because $\tilde u(0)=0$. Thus, ${\partial u\over \partial y}(z)\geq 0$ for $z\in L$.

Assume now that ${\partial u\over \partial y}(z)\geq 0$ for $z\in L$. We have to show that $\tilde u$ is subharmonic. 
Recall that if a sequence of subharmonic functions converges locally uniformly, then the limit is also a subharmonic function. 
Therefore, we can replace $u$ by $u+\epsilon y$ for $\epsilon>0$ in order to assume that ${\partial u\over \partial y}(z)> 0$ for $z\in L$.
By continuity, the last inequality holds for $z$ in a neighbourhood of $L$.

 The problem is local near the points of of $L$. So without loss of generality, we can assume that $\Omega$ is the square 
$(-1,1)\times (-1,1)$ and that ${\partial u\over \partial y}(z)>0$ on $\Omega_+\cup L$. It follows that the restriction of $u$ to $\{x\}\times [0,1)$ is strictly increasing for every $x\in (-1,1)$.
For  $0<\eta<1/2$, consider the functions 
$$v_\eta (z) := u(2\eta i +\overline z) \qquad \text{and} \qquad \tilde u_\eta(z):=
\begin{cases}
 u(z) &  \text{when } \ \eta\leq y <1\\
 v_\eta(z) & \text{when } \ -1+2\eta<y\leq \eta.
\end{cases}
$$
Using that $u$ is increasing in vertical lines, we deduce that 
$$ \qquad \tilde u_\eta(z):=
\begin{cases}
 u(z) &  \text{when } \ \eta < y <1\\
 \max(u(z),v_\eta(z))  & \text{when } \ 0< y <2\eta \\
 v_\eta(z) & \text{when } \ -1+2\eta<y <\eta.
\end{cases}
$$

Observe that $v_\eta$ is subharmonic on $(-1,1)\times (-1+2\eta, 2\eta)$.
Therefore, the last formula for $\tilde u_\eta$ implies that this function is subharmonic everywhere in 
$(-1,1)\times (1-2\eta,1)$. We use that the maximum of two subharmonic functions is subharmonic. Finally,
when $\eta\to 0$ we see that $\tilde u_\eta\to \tilde u$ locally uniformly on $(-1,1)\times (-1,1)$. 
It follows that $\tilde u$ is subharmonic in $(-1,1)\times (-1,1)$. 

\medskip

(2) Recall that $\ddc\tilde u$ is a  positive measure on $\Omega_+\cup\Omega_-$. So the problem is local in a neighbourhood of each point of $L$. We can assume as above that $\Omega$ is the square $(-1,1)\times (-1,1)$ and that the function 
$m(x):={1\over \pi} {\partial u\over \partial y}(x)$ is uniformly continuous and bounded on $L=(-1,1)$.  
We extend $m(x)$ to a function on $\R$ which vanishes outside $L$. Define $\nu:=m(x)dx$ which is a finite measure on $\C$ with support in $\overline L$. We first show that $\ddc \tilde u\geq \nu$.
This implies that $\ddc\tilde u$ is a measure on $\Omega.$

Consider the logarithmic potential of the measure $\nu$  defined by 
$$v(z):= \int_\R \log|z-t| m(t)dt \quad \text{for } z\in\C.$$
It satisfies $v(\overline z)=v(z)$, $\ddc v=\nu$ and therefore, $v$ is harmonic on $\C\setminus \overline L$. It is not difficult to show that this function is continuous in $\C$. Moreover,
for $z=x+iy$ with $y>0$, $x':=y^{-1}x$ and $t':=y^{-1}(t-x)$, we have
$${\partial v\over\partial y}(z) = {\partial \over\partial y} \int_\R {1\over 2} \log ((x-t)^2+y^2) m(t) dt=\int_\R {y\over (x-t)^2+y^2} m(t) dt=\int_\R {m(x+yt') \over t'^2+1}  dt'.$$ 
Recall that the integral of $(t'^2+1)^{-1}$ on $\R$ is $\pi$. Thus, it is not difficult to see from the last computation that the function ${\partial v\over\partial y}(z)$ on $\Omega_+$ extends to a continuous function on $\Omega_+\cup L$. It converges uniformly to $\pi m(x)$ when $y\to 0$.  We conclude that ${\partial v\over\partial y}(z)$ exists and is continuous on $\Omega_+\cup L$ and equal to ${\partial u\over\partial y}(z)$ on $L$.

Since the function $v$ is harmonic on $\Omega_+$, the function $u-v$ is subharmonic on $\Omega_+$. We can apply the first assertion (1) to $u-v, \tilde u-v$ instead of $u,\tilde u$ and deduce that $\tilde u-v$ is subharmonic. It follows that $\ddc \tilde u\geq\ddc v=\nu$. For the rest of the proof, without loss of generality, we can replace $u,\tilde u$ by $u-v,\tilde u-v$ in order to assume that ${\partial u\over\partial y}(z)=0$ on $L$ and we still need to check that $\ddc\tilde u$ has no mass on $L$.

Assume by contradiction that $\ddc\tilde u$ doesn't vanish on $L$. Replacing $u, \tilde u$ by $u^\epsilon,\tilde u^\epsilon$ which are defined at the beginning of the proof, we can assume that the restriction of $\ddc \tilde u$ to $L$ is a positive measure absolutely continuous with continuous and bounded density with respect to the Lebesgue measure on $L$. 
Denote this measure by $\nu'=m'(x)dx$ and $v'(z)$ its logarithmic potential. Since $\ddc \tilde u\geq\nu'$, the function $\tilde u-v'$ is subharmonic in $\Omega$. A computation as before shows that 
$${\partial (u-v')\over \partial y}(z) =-{\partial v'\over \partial y}(z) =-\pi m'(z) \quad \text{for}\quad z\in L.$$ 
This contradicts the assertion (1) applied to $u-v'$ and $\tilde u-v'$ which implies that the last partial derivative of $u-v'$ should be non-negative on $L$. This ends the proof of the lemma.
\endproof

Now, we recall some notions and results related to the Melrose-Zworski distribution. 
Let $\rho$ be  the continuous  function   on $\overline{\C}_+\setminus \{0\}$ defined by
\begin{equation}\label{e:rho}
 \rho(z):=\log{1+ \sqrt{1-z^2}\over   z} - \sqrt{1-z^2}
\end{equation}
which extends the  real-valued function in $z\in (0,1)$  given by the  same  formula.
Set
\begin{equation}\label{e:h_d}
 h_d(\theta):={4\over  (d-2)!}\int_0^\infty { \max\big(-\Re \rho(te^{i\theta}),0\big)\over  t^{d+1}}dt \quad \text{for}\quad 0<\theta<\pi,
\end{equation}
and set $h_d(0):=0$ and $h_d(\pi):=0.$ It can be  shown  that  $h_d$ is continuous  on $[0,\pi]$ and  is  a real analytic function on $(0,\pi).$
The constant $c_d$ appearing in the Introduction  is given by
\begin{equation}\label{e:c_d}
 c_d:= {d\over 2\pi} \int_0^\pi h_d(\theta)d\theta= {2d\over  \pi(d-2)!}\int_{\Im z>0} { \max\big(-\Re \rho(z),0\big)\over  |z|^{d+2}}dxdy.
\end{equation}
We infer from     \eqref{e:rho} and  \eqref{e:h_d}   that $\Re\rho(z)$ is invariant under the map $z\mapsto -\overline z$ and hence
\begin{equation}
 \label{e:h_symmetry}
 h_d(\pi/2+\theta)=h_d(\pi/2-\theta) \quad\text{for}\quad  0\leq \theta\leq \pi/2.  
\end{equation}
Using  the gamma  function $\Gamma$ consider also  the  constant  
\begin{equation}\label{e:e_d}
 e_d:=\sqrt{\pi}{\Gamma( {d-1\over 2} )\over  (d-2)!  \Gamma(1+d/2)}\cdot 
\end{equation}

We  know by  Christiansen \cite[Lemma 3.3]{Christiansen12} that

\begin{lemma}\label{l:hd}
The function  $h_d(\theta)$ is of class $\Cc^1$ on $[0,\pi].$  Moreover, we have
\begin{equation*}
h'_d(0+)=\lim_{\theta\to 0+} h'_d(\theta)= e_d \quad \text{and} \quad  h'_d(\pi-0)=\lim_{\theta\to \pi-0} h'_d(\theta)=-e_d.
\end{equation*}
Here, $h'_d(0+)$, $h'_d(\pi-0)$ denote respectively the right and left derivatives of $h_d$ at $0$ and $\pi$.
\end{lemma}

The measures involved in our main results are supported 
 by the lower half-plane. However, it is more convenient to work with measures which are symmetric with respect to the real line because their potentials are easier to compute, see for instance Lemma \ref{l:subharmonic}.
This is the reason why we introduce the following notions. 

First, we extend the function $\kappa(z)$ defined in the Introduction on $\C_-$ to $\C\setminus \R$ using the equation $\kappa(z)=\kappa(\overline z)$.  Define

\begin{equation}\label{e:H_Z}
H_Z(z):= \begin{cases}
          0 & \text{for}\ z\in\C_+\\
c_d^{-1}|z|^d h_d(|\theta|) & \text{for}\ z\in\overline{\C}_-
\end{cases}
\end{equation}
and
\begin{equation}\label{e:H}
        H(z):=c_d^{-1}|z|^d h_d(|\theta|) =H_Z(z)+H_Z(\overline z)
=\begin{cases}H_Z(\overline z) & \text{for}\quad  z\in\overline\C_+\\
                H_Z(z) & \text{for}\quad  z\in\overline \C_-,
              \end{cases}
\end{equation}  
where $-\pi\leq \theta\leq \pi$ is the argument of $z$.

\begin{lemma} \label{l:ddcH}
We have 
$$\ddc H(z)={i\over 2} \kappa(z) dz\wedge d\overline z \quad \text{on} \quad \C\setminus\R.$$
Moreover, $\kappa(z)$ is an analytic real function which satisfies
$$\kappa(tz)=t^{d-2}\kappa(z)\qquad\text{and}\qquad  \kappa(z)=O(|y|^{1/2}|z|^{d-5/2}) \quad \text{as } z\to\infty.$$
In particular,  when $|z|$ is bounded, $\kappa(z)$ is bounded and $\kappa(z)=O(|y|^{1/2})$ as $y$ tends to $0$.
\end{lemma}
\proof
Note that ${i\over 2} dz\wedge d\overline z=dx\wedge dy$ is the area form associated to the Lebesgue measure on $\C$. 
Observe that since $\kappa(z)$ is invariant by the maps $z\mapsto\pm\overline z$, it is enough to consider the case where $0<\theta\leq \pi/2$. 
Recall that 
$$\Delta ={\partial^2 \over \partial r^2} +{1\over r} {\partial \over \partial r}+{1\over r^2} {\partial^2 \over \partial \theta^2}
\qquad \text{and} \qquad H(z)=c_d^{-1}r^d h_d(\theta) \quad \text{for} \quad z\in\C_+.$$
Therefore, we have using the definition of $\kappa(z)$ 
$$\ddc H(z)={i\over\pi}\ddbar H(z)= {i\over 4\pi} \Delta H(z) dz\wedge d\overline z={i\over 2} \kappa(z) dz\wedge d\overline z.$$
This gives us the first identity.

The second identity is a direct consequence of the definition of $\kappa(z)$. 
We prove now the next estimate in the lemma using the previous ones and will see in the proof that $\kappa(z)$ is analytic real.
Recall from \cite{Stefanov} that the set $\Sigma:=\{z\in \C_+: \ \Re\rho(z)=0\}$ is a smooth analytic real curve intercepting the real line $\R$ at two points 1 and $-1$. In polar coordinates $(r,\theta)$, it is given by an equation $r=r_0(\theta)$ with $0<\theta<\pi$, where $r_0(\theta)$ can be extended to an analytic real function in a neighbourhood of $[0,\pi]$. For $z=re^{i\theta}$, we have $\Re\rho(z)<0$ is and only if $r>r_0(\theta)$. Moreover, we have $r_0(\theta)>1/2$ for $\theta\in [0,\pi]$.
Define $s_0(z):=r_0(\theta)|z|^{-1}$. Using  \eqref{e:h_d}, \eqref{e:H_Z} and the variable $s:=|z|^{-1}t$, we have 
$$H(z) = {4\over c_d (d-2)!} \int_0^\infty {\max (-\Re\rho(sz),0)\over s^{d+1}} ds
= -{4\over c_d (d-2)!} \Re \int_{s_0(z)}^\infty {\rho(sz)\over s^{d+1}} ds .$$

Using the first identity and the fact that $\partial^2/\partial z\partial\overline z$ is a real operator, we have 
$$\kappa(z)=2{\partial^2 H(z)\over \partial z\partial\overline z}= -{8\over c_d (d-2)!} \Re {\partial^2\over \partial z\partial\overline z} \int_{s_0(z)}^\infty {\rho(sz)\over s^{d+1}} ds.$$
It is clear now that $\kappa(z)$ is a real analytic function. We continue the proof of the last identity in the lemma.
By the second identity in the lemma, it is enough to show that $\kappa(z)=O(y^{1/2})$ when $|z|=1$ and $z\to 1$, or equivalently, the last second order derivative satisfies the same property.

Now, since $\rho(z)$ is holomorphic, its partial derivative in $\overline z$ vanishes. We deduce that
$${\partial^2\over \partial z\partial\overline z} \int_{s_0(z)}^\infty {\rho(sz)\over s^{d+1}} ds 
=-{\partial\over \partial z} \Big[{\rho(s_0(z)z)\over s_0(z)^{d+1}}{\partial s_0(z)\over \partial\overline z}\Big]\cdot
$$
Recall that $s_0(z)z$ belongs to the curve $\Sigma$ and when $z\to 1$ we also have $s_0(z)z\to 1$. Since $r_0(\theta)$ is analytic in a neighbourhood of $[0,\pi]$, it is enough to check that $|\rho(z)|+|\rho'(z)|=O(|z-1|^{1/2})$ when  $z\to 1$.  This property is clear because by \eqref{e:rho}, for $z\to 1$, we have 
$$\rho(z)=-\log z -{1\over 2} (1-z^2) + O(|1-z^2|^{3/2}) \qquad \text{and} \qquad
\rho'(z)=O(|1-z^2|^{1/2}).$$
The lemma follows.
\endproof

\begin{lemma} \label{l:mu}
The functions $H_Z, H$ are subharmonic on $\C$ and $\mu^-,\mu_\MZ, \ddc H$ are positive measures on $\C$. 
Define the positive measure $\mu$ on $\C$ by $\mu:=\ddc H$. Let $\mu^+$ be the image of $\mu^-$ by the map $z\mapsto \overline z$. Then we have
$$\ddc H_Z=\mu^-+\mu^0=\mu_\MZ  \qquad \text{and} \qquad \mu=\mu^++\mu^-+2\mu^0.$$  
\end{lemma}
\proof
For the first assertion, it is enough to show that $H_Z(z)$ is subharmonic because this property implies that $H_Z(\overline z)$ and hence $H(z)$ are also subharmonic. Using properties of $h_d$ and Lemma \ref{l:hd}, we see that $H_Z(z)$ is continuous, non-negative on $\overline\C_+$ and vanishes on $\R$. By Lemma \ref{l:u0}, it is enough to check that $H_Z(z)$ is subharmonic on $\C_+$. 

Using  \eqref{e:h_d}, \eqref{e:H_Z} and the variable $s:=|z|^{-1}t$, we have for $z\in\C_+$
$$H_Z(z) = {4\over c_d (d-2)!} \int_0^\infty {\max (-\Re\rho(sz),0)\over s^{d+1}} ds.$$
Since the function $z\mapsto \rho(sz)$ is holomorphic, the function $z\mapsto -\Re\rho(sz)$ is harmonic and the function 
$z\mapsto \max (-\Re\rho(sz),0)$ is subharmonic. 
We easily deduce that $H_Z(z)$ is subharmonic and hence $H$ is also subharmonic.

We prove now the two identities in the lemma.  
Since $H_Z$ vanishes on $\C_+$, the measure $\mu_\MZ$ also vanishes there. It follows from 
Lemma \ref{l:ddcH} that $\ddc H_Z=\mu^-$ on $\C_-$. If $m$ denotes the restriction of $\ddc H_Z$ to $\R$, then we have $\ddc H_Z=\mu^-+m$. We also deduce from \eqref{e:H} that $\mu=\mu^++\mu^-+2m$. 
So $2m$ is the restriction of $\mu=\ddc H$ to $\R$ and it remains to check that $m=\mu^0$.

By Lemma \ref{l:hd} and \eqref{e:H}, the function $H$ is $\Cc^1$ on $\overline \C_+$.  Moreover, on $\overline\C_+$, we have for $x> 0$
$${\partial H\over\partial y} (x) = c_d^{-1} x^d x^{-1}h_d'(0)=c_d^{-1} x^{d-1}e_d$$
and for $x<0$
$${\partial H\over\partial y} (x) = c_d^{-1} |x|^d x^{-1}h_d'(\pi-0)=c_d^{-1} |x|^{d-1}e_d.$$
It is also easy to check that this derivative vanishes at $0$. Now, the result follows from the second assertion of Lemma \ref{l:subharmonic} and the definition of $\mu^0$ in the Introduction.
\endproof

 \begin{lemma}\label{l:mass-theta1-theta2}
For $0\leq \theta_1<\theta_2\leq \pi,$ let $\Omega(\theta_1,\theta_2)$ be as in Corollary \ref{C:Christiansen}. Then 
 we have 
  \begin{equation*}
   \mu_\MZ\big(\Omega(\theta_1,\theta_2)\big)=   {1\over 2\pi dc_d}\Big[ h'_d(\theta_2)-h'_d(\theta_1)  +d^2\int_{\theta_1}^{\theta_2} h_d(\theta)d\theta  \Big],
\end{equation*}
where we replace $h_d'(\theta_1)$ and $h_d'(\theta_2)$ by $h_d'(0+)$  and $h_d'(\pi-0)$ respectively when $\theta_1=0$ and $\theta_2=\pi$. Moreover, we have 
$$\mu_\MZ([0,1])=\mu_\MZ([-1,0])= {e_d\over 2\pi d c_d}    \qquad \text{and} \qquad \mu_\MZ(\D)=\mu_\MZ(\overline\D_-)=1.$$
\end{lemma}
 \proof
We have 
\begin{eqnarray*}
\mu_\MZ\big(\Omega(\theta_1,\theta_2)\big) &=& \int_{0<r<1, \theta_1<\theta<\theta_2} \kappa(z) dxdy \\
& = & \int_{0<r<1, \theta_1<\theta<\theta_2}  {1\over 2\pi c_d} r^{d-2} \big[d^2h_d(\theta) +h''_d(\theta)\big]  rdrd\theta.
\end{eqnarray*} 
The first identity in the lemma follows easily.

For the second assertion in the lemma, we have 
$$\mu_\MZ([0,1])=\mu^0([0,1])=\int_0^1 {e_d\over 2\pi c_d} x^{d-1} dx={e_d\over 2\pi dc_d}\cdot$$
Similar identities also hold for $[-1,0]$. 
Recall that $\mu_\MZ$ has no mass on $\C_+$ and is absolutely continuous with respect to the Lebesgue measure on $\C_-$. Therefore, we have 
$$\mu_\MZ(\D)=\mu_\MZ(\overline\D_-)=\mu^-(\D_-)+\mu^0([-1,1])=\mu^-(\D_-)+{e_d\over \pi dc_d} \cdot$$ 
Using the first identity in the lemma, \eqref{e:c_d} and Lemma \ref{l:hd}, we get 
$$\mu_\MZ(\D_-)=\mu^{-}(\D_-)=  {1\over 2\pi dc_d}\big[ -2e_d  +2\pi d c_d  \big]=1-{e_d\over \pi dc_d}\cdot$$
It follows that $\mu_\MZ(\D)=\mu_\MZ(\overline\D_-)=1$.
This completes the proof of the lemma.

\section{Distribution of the resonances} \label{s:resonances} 

In this section, we will prove the main results stated in the Introduction. Recall that in order to simplify the proof we will symmetrize the measures and potentials with respect to the real axis. 
Let $S_V(z)$ be  the     scattering matrix associated to   $R_V(z)$   and $s_V(z):=\det S_V(z)$, see e.g.  \cite{DZ}.
Recall that $s_V(z) s_V(-z)=1$ and
if $s_V$   has  poles in the  closed upper half-plane $\overline{\C}_+,$ it has  only  finitely many.
Moreover, with  at  most  finitely  many  exceptions, the zeros of  $s_V(z)$  coincide  with  the poles of   $R_V(-z),$  and  the multiplicities  agree.  Therefore, our study uses in a crucial way the function $s_V(z)$.
The following result was obtained in  \cite{DinhVu}, see also \cite[Theorem 5]{Stefanov}.   

\begin{proposition}\label{P:Dinh-Vu}
There is a constant $A>0$  depending only on  $d,a$ and $V$ such that 
$$\log |s_V(re^{i\theta})| \le h_d(\theta)a^d r^d+ Ar^{d-1}\log r$$
for all $r\geq A$  and  $\theta\in [0,\pi]$.
\end{proposition}

We also have the following estimate.

\begin{proposition}\label{P:Christiansen}
We have  for $r\geq 0$ 
$$ \Big|N_V(r)- \frac{1}{2\pi} \int_0^{\pi} \log |s_{V}(r e^{i\theta})| d\theta\Big|
\leq Ar^{d-1}+A,$$
where $A>0$ is a constant depending only on $d,a$ and $V$.
\end{proposition}
\proof
By  Christiansen \cite[(3.2)]{Christiansen12} and Stefanov  \cite[Lemma 2]{Stefanov}, the estimate holds for $r$ large enough, see also Froese \cite{Froese2}, Petkov-Zworski \cite{PZ}.
Since $s_V(z) s_V(-z)=1$, we have $s_V(0)\not=0$ and therefore, the estimate holds for $r$ small enough. Finally, when $r$ is bounded by two positive constants, since $s_V$ is a meromorphic function, the integral in the lemma is bounded. We easily deduce the proposition by taking $A$ large enough.
\endproof

The  following  estimate  due  to Christiansen \cite[Lemma 3.1]{Christiansen12}  will be  needed.

\begin{lemma}\label{l:Christiansen} 
There is  a  constant $A$ depending only on $d,a$ and $V$  such that
 $$ \big|  {d\over dz }\log s_V(z)   \big|\leq  A |z|^{d-2} $$
 for  $z\in \R$ with $|z|$ large enough.
\end{lemma}

Finally,  we  will  also use the  following result  which  relates the asymptotic  behavior of $n_V(r)$ and $N_V(r)$. It was obtained in 
\cite[Lemma 1]{Stefanov} and  \cite[Prop. 5.1]{DinhVu}.

\begin{lemma}   \label{l:nVNV} 
Let $\nu$ and $A$  be   two constants    such that $0<\nu<d$ and  $A>0.$ Then the following holds.
\begin{enumerate}
\item $n_V(r)=Ar^d +o(r^d)$ as $r\to\infty$     if and  only if $N_V(r)={Ar^d\over d}  +o(r^d)$ as $r\to\infty.$     
\item    If $n_V(r)=Ar^d+O(r^{d-\nu})$  as $r\to\infty,$ then $N_V(r)={Ar^d\over d}+O(r^{d-\nu})$  as $r\to\infty.$
\item    If $N_V(r)=Ar^d+O(r^{d-\nu})$  as $r\to\infty,$ then $n_V(r)={Ar^d\over d}+O(r^{d-\nu/2})$  as $r\to\infty.$
\end{enumerate}
\end{lemma}

For  $r>0$, consider the following function
\begin{equation}\label{e:uVr}
u_{V,r}(z):=
\begin{cases}
{1\over  c_d a^d r^d}\log |s_V(-r\overline z)| & \text{for} \quad z\in\overline\C_+,\\
   {1\over  c_d a^d r^d}\log |s_V(- r z)|&  \text{for} \quad  z\in\overline\C_-.
  \end{cases}
 \end{equation}

\begin{proposition} \label{p:uVr}
 The following properties hold.
 \begin{itemize}
 \item[(1)]  If $V$ belongs to the Christiansen class $\fM_a$, then 
$$\sup_{s\geq 1}  s^{-d-2} \|u_{V,r} -H\|_{L^1(\D(s))}\to  0 \quad\text{as}\quad  r\to\infty; $$
 \item[(2)] If $V$ belongs  to the  class $\fM^\nu_a$ for some $0<\nu \leq 1,$ then for every $\eta>0$
$$  \sup_{s \geq 1} s^{-d-2} \|u_{V,r} -H\|_{L^1(\D(s))}  =O(r^{-\nu+\eta})  \quad\text{as}\quad  r\to\infty.$$
\end{itemize}
 \end{proposition}
\proof
(1)  It is enough to consider $r$ large enough. Define 
$$ \eta(r):={1\over c_d a^d r^d}\Big|N_V(r)-{c_da^dr^d\over d}\Big|. $$
Since  $V$ is  a function in the Christiansen class $\fM_a$, by the first assertion of Lemma \ref{l:nVNV}, we have that $\eta(r)$ tends to 0 when $r$ tends to infinity.  
By Proposition \ref{P:Christiansen}, we have
$$\Big| {1\over2\pi} \int_0^\pi \log |s_V(re^{i\theta})|d\theta-{c_da^dr^d\over d}\Big|\leq c_da^d r^d\eta(r)+Ar^{d-1}+A$$
for $r\geq 0$. 
Using \eqref{e:uVr}, we obtain that 
$$\Big| {1\over2\pi} \int_0^\pi u_{V,r}(te^{i\theta})d\theta-{t^d\over d}\Big| \leq t^d\eta(tr) +A't^{d-1}r^{-1}+A'r^{-d},$$
for some constant $A'>0$. Therefore, by  \eqref{e:c_d}, we can rewrite the  last line as
$$\Big| {1\over2\pi} \int_0^\pi \Big[u_{V,r}(te^{i\theta}) -{ h_d(\theta)t^d\over    c_d}\Big]d\theta\Big|  \leq t^d\eta(tr) +A't^{d-1}r^{-1}+A'r^{-d},$$
or equivalently, by \eqref{e:H}, 
$$\Big| {1\over2\pi} \int_0^\pi \Big[u_{V,r}(te^{i\theta}) -H(te^{i\theta})\Big]d\theta\Big|  \leq t^d\eta(tr) +A't^{d-1}r^{-1}+A'r^{-d}.$$

Finally, using the polar coordinates and Fubini's theorem, we deduce that
$$\|u_{V,r} -H\|_{L^1(\D(s))}\leq \int_0^s \big[t^d\eta(tr) +A't^{d-1}r^{-1}+A'r^{-d}\big] tdt \leq \int_0^s t^{d+1}\eta(tr)dt + A''s^{d+1}r^{-1}$$
for some constant $A''>0$.
By a simple change of variable, we see that the last integral is equal to $s^{d+2}\gamma(rs)$ with
$$\gamma(rs):= (rs)^{-d-2}\int_0^{rs} t^{d+1}\eta(t) dt.$$
Since $\eta(t)$ tends to 0 as $t$ tends to infinity, $\gamma(rs)$ converges to 0 as $rs$ tends to infinity. In particular, $\gamma(rs)$ tends to 0 when $r$ tends to infinity, uniformly in $s$ because $s\geq 1$. This completes the proof of the first assertion of the proposition.

\medskip
(2)  Assume  now that   $V$ is in  $\fM^\nu_a$. We can also assume that $\eta<\nu$.  By  the second assertion of Lemma \ref{l:nVNV},  we have 
$\eta(rs)\leq A_\eta (rs)^{-\nu+\eta}$ for some constant $A_\eta>0$. 
It follows that 
$$\gamma(rs)\leq (rs)^{-d-2} A_\eta (rs)^{d+2-\nu+\eta}\leq A_\eta r^{-\nu+\eta}$$
since $s\geq 1$. The proposition follows.
\endproof

Recall  that $s_V(z)s_V(-z)=1$ for $z\in\C$ and 
$s_V$   has  only a finite number of poles in the  closed upper half-plane $\overline{\C}_+$. We denote them by $z_1,\ldots,z_m,$  where each pole is  repeated according to its multiplicity.  It is convenient to modify the functions $s_V$ and $u_{V,r}$ slightly.
Define 
\begin{equation}\label{e:sVhat}
 \hat s_V(z):=\prod_{j=1}^m  {z-z_j\over z+z_j}s_V(z).
\end{equation}
We  see that  $\hat s_V$ is   a holomorphic  function on an open neighborhood of  $\overline{\C}_+$   with neither  zeros  nor poles on the  real line $\R.$ Define also  
\begin{equation}\label{e:uVrhat}
\hat u_{V,r}(z):=
\begin{cases}
{1\over c_d a^d r^d}\log |\hat s_V(-r\overline z)| & \text{for} \quad z\in \overline \C_+\\
 {1\over c_d a^d r^d}\log |\hat s_V(-rz)| &  \text{for} \quad  z\in\overline \C_-.
\end{cases}
\end{equation}

\begin{lemma} \label{l:sVhat}
There is a constant $A>0$ such that 
$$\Big|{d\over dz} \log \hat s_V(z) \Big| \leq A (|z|^{d-2}+1) \quad \text{for} \quad z\in\R.$$
\end{lemma}
\proof
By Lemma \ref{l:Christiansen} and \eqref{e:sVhat}, there are constants $A>0$ and $R>0$ such that the above inequality holds for $z\in\R$ such that $|z|\geq R$. On the other hand, $\hat s_V$ is holomorphic and does not vanish in a neighbourhood of $\R$. Therefore, the same inequality holds for $z\in [-R,R]$ provided that $A$ is large enough. The lemma follows.
\endproof

\begin{lemma} \label{l:uVrhat}
 We have the following properties.
 \begin{itemize}
 \item[(1)]  If $V$ belongs to the Christiansen class $\fM_a$, then 
$$\sup_{s\geq 1}  s^{-d-2} \|\hat u_{V,r} -H\|_{L^1(\D(s))}\to  0 \quad\text{as}\quad  r\to\infty; $$
 \item[(2)] If $V$ belongs  to the  class $\fM^\nu_a$ for some $0<\nu \leq 1,$ then for every $\eta>0,$
$$\sup_{s \geq 1} s^{-d-2} \|\hat u_{V,r} -H\|_{L^1(\D(s))}  =O(r^{-\nu+\eta})  \quad\text{as}\quad  r\to\infty.$$
\end{itemize}
 \end{lemma}
\proof
By \eqref{e:uVr} and \eqref{e:uVrhat}, we have 
\begin{eqnarray*}
\hat u_{V,r}(z)- u_{V,r}(z) &  = & {1\over c_d a^d r^d}\sum_{j=1}^m  \log|rz-z_j|-\log|rz + z_j| \\
& = &  {1\over c_d a^d r^d}\sum_{j=1}^m  \log|z -r^{-1} z_j|-\log|z+r^{-1} z_j|.
\end{eqnarray*}
So for $r$ large enough, we have $|r^{-1} z_j|\leq 1$ and hence the $L^1$-norm of $ \log|z\pm r^{-1} z_j|$ on $\D(s)$ is bounded by the $L^1$-norm of $\log|z|$ on $\D(s+1)$. The last one is bounded by 
a constant times $s^2\log(s+1) $. Therefore, the lemma follows from the last identities and Proposition \ref{p:uVr}.
\endproof

Now, define $\hat \mu_{V,r} :=\ddc \hat u_{V,r}$. 
Since $\hat s_V$ has no pole on $\overline\C_+$ and no zero on $\R$, the function $\hat u_{V,r}$ is subharmonic on $\C_+\cup \C_-$ and is symmetric with respect to $\R$. We can apply
Lemma \ref{l:subharmonic} to this function on a suitable neighbourhood $\Omega$ of $\R$  and deduce that $\hat\mu_{V,r}$ is a measure on $\C$. 
Denote also by $\hat\mu_{V,r}^+, \hat\mu_{V,r}^-$ and $2\hat\mu_{V,r}^0$ the restrictions of $\hat\mu_{V,r}$ to $\C_+,\C_-$ and $\R$, respectively. Let
$\fZ_V$ denote the set of zeros of $s_V(z)$ on $\C_+$ where each point is repeated according to its multiplicity. 
This is also the zero set of $\hat s_V(z)$ on $\C_+$. We see that 
$$\hat \mu_{V,r}^+ = {1\over c_d a^d r^d} \sum_{z\in \fZ_V} \delta_{z/r}.$$
Moreover, by definition of $\hat u_{V,r}$, $\hat\mu_{V,r}^-$ is the image of $\hat\mu_{V,r}^+$ by the map $z\mapsto \overline z$. 
So both $\hat\mu_{V,r}^+$ and  $\hat\mu_{V,r}^-$ are positive measures.

\begin{lemma} \label{l:muVrhat}
Let $0<\gamma\leq 2$ be a constant and let $\Omega, \Omega'$ be two open discs in $\C$ centered at $0$ with $\Omega\Subset\Omega'$. Then there is a constant $A>0$ independent of $r$ such that $Ar^{-1}\Leb_\R\pm \hat\mu_{V,r}^0$ are positive measures on $\Omega$ and
$$ \dist_{\Omega,\gamma}\big(\hat\mu_{V,r}^++\hat\mu_{V,r}^-, \mu\big)\leq Ar^{-1}+A\|\hat u_{V,r}-H\|_{L^1(\Omega')}^{\gamma/2}.$$
In particular, the mass of $\hat\mu_{V,r}^0$ on $\Omega$ tends to $0$ as $r$ tends to infinity.
\end{lemma}
\proof
It is clear that the last assertion is a consequence of the first one.
Write $\Omega=\D(s)$ and $\Omega'=\D(s')$ with $s<s'$. 
Since $\hat u_{V,r}$ is symmetric with respect to $\R$, we can apply Lemma \ref{l:subharmonic} to this function. 
Denote by $u$ the restriction of $\hat u_{V,r}$ to $\D(s')\cap \overline\C_+$ as in Lemma \ref{l:subharmonic}.
By Lemma \ref{l:sVhat}, we have $\big|{\partial u \over\partial y}\big|\leq Ar^{-1}$ on $\D(s')\cap \R$ for some constant $A>0$ independent of $r$. Then, using Lemma \ref{l:subharmonic}, we see that $Ar^{-1}\Leb_\R\pm \hat\mu_{V,r}^0$ are positive measures on $\D(s')$. The first assertion of the lemma follows.

For the estimate in the lemma, by \eqref{e:dist-gamma-gamma}, we only need to check that 
$$\dist_{\D(s'),2}\big(\hat\mu_{V,r}^++\hat\mu_{V,r}^-, \mu\big)\leq Ar^{-1}+A\|\hat u_{V,r}-H\|_{L^1(\D(s'))}$$
or equivalently if $\varphi$ is a $\Cc^2$ function with compact support in $\D(s')$ with $\|\varphi\|_{\Cc^2}\leq 1$, we need to show that 
$$\big|\langle \hat\mu_{V,r}^++\hat\mu_{V,r}^- - \mu, \varphi\rangle \big| \leq Ar^{-1}+A\|\hat u_{V,r}-H\|_{L^1(\D(s'))}.$$

Consider the function $u^*(z):=\hat u_{V,r}(z) + 2\pi A r^{-1}|y|$. It follows from the above arguments that  $\mu^*:=\ddc u^*- (\hat\mu_{V,r}^++\hat\mu_{V,r}^-)$ is a positive measure on $\D(s')$ of mass bounded by a constant times $r^{-1}$. Therefore, since $\mu=\ddc H$ and $\ddc\varphi$ is bounded, we have 
\begin{eqnarray*}
\big|\langle \hat\mu_{V,r}^++\hat\mu_{V,r}^- - \mu, \varphi\rangle \big|  & \leq & \big|\langle \mu^*, \varphi\rangle \big| +
\big|\langle \ddc u^* -\ddc H, \varphi\rangle \big| \\
& \lesssim &  r^{-1} + \big|\langle u^* -H, \ddc \varphi\rangle \big| \\
& \lesssim &  r^{-1} + \| u^* -H\|_{L^1(\D(s'))} \\
& \lesssim &  r^{-1} + \| \hat u_{V,r} -H\|_{L^1(\D(s'))}.
\end{eqnarray*}
So the estimate in the lemma holds for some constant $A>0$ large enough. This completes the proof of the lemma.
\endproof

\medskip

\noindent{\bf End of the proof of  Theorem \ref{T:main_1}.}
By Lemmas \ref{l:uVrhat} and \ref{l:muVrhat}, $\dist_{\Omega,2}(\hat\mu_{V,r}^++\hat\mu_{V,r}^-,\mu)$ tends to 0 as $r$ tends to infinity for any disc $\Omega$ of center 0. It follows that
$\hat\mu_{V,r}^++\hat\mu_{V,r}^-$ tends to $\mu$ weakly. Recall that $\hat\mu_{V,r}^+$ and $\hat\mu_{V,r}^-$ have supports in $\C_+$ and $\C_-$ respectively. 
Moreover, they are symmetric each to other with respect to $\R$. Recall also from Lemma \ref{l:mu} that $\mu=\mu^++\mu^-+2\mu^0$ where $\mu^+, \mu^-,2\mu^0$ are the restrictions of $\mu$ to $\C_+,\C_-$ and $\R$ respectively. Therefore, the fact that $\hat\mu_{V,r}^++\hat\mu_{V,r}^-$ tends to $\mu$ implies that $\hat\mu_{V,r}^-$ converges weakly to $\mu^-+\mu^0$. The last measure is equal to $\mu_\MZ$.

Recall that with at most finitely many of exceptions, a point $z$ is a resonance, i.e. a pole of $R_V$, if and only if $-z$ is a zero of $\hat s_V$ and their multiplicities agree. Therefore, the measure $\mu^\fR_{V,r}-\hat\mu^-_{V,r}$ is a finite combination of atoms with total mass bounded by a constant times $r^{-d}$. We deduce that $\mu^\fR_{V,r}$ and $\hat\mu^-_{V,r}$ have the same limit $\mu_\MZ$ when $r$ tends to infinity. This is the first assertion in the theorem. 

Now, consider the second assertion in the theorem. Since there are finitely many of resonances in $\overline\C_+$, it is enough to prove the statement for $W=\Omega$ and for $W=\overline\Omega$.
Since  $n_{V,W}(r) = c_d a^dr^d \mu_{V,r}^\fR(W)$, 
we easily deduce the second assertion in the theorem from the first one by using Lemma \ref{l:simple_cv}. 
\hfill  $\square$

\medskip

\noindent{\bf End of the proof of  Corollary \ref{C:Christiansen}.} 
Consider first the case where $0<\theta_1<\theta_2<\pi$. 
By Theorem \ref{T:main_1} applied to $\Omega:=\Omega(\theta_1,\theta_2)$, we have 
$$n_{V,\Omega}(r)= c_da^d\mu_\MZ(\Omega)r^d+o(r^d).$$
We easily deduce the result from Lemma \ref{l:mass-theta1-theta2}.

Consider now the case where $\theta_1=0$ and $\theta_2<\pi$. The other cases can be treated in the same way.
Let $\Omega^*$ be the open sector of the unit disc which is the union of $\Omega$, its symmetry with respect to $\R$ and the radius $(-1,0)$. Applying Theorem \ref{T:main_1} to $\Omega^*$ instead of $\Omega$ and using that $\mu_\MZ(\C_+)=0$, we obtain
$$n_{V,\Omega^*}(r)= c_da^d\big[\mu_\MZ((-1,0))+ \mu_\MZ(\Omega)\big]r^d+o(r^d).$$
Since there are only finitely many of resonances in $\overline\C_+$, the last identity still holds if we replace $\Omega^*$ by $\Omega$. 
Lemmas \ref{l:hd} and  \ref{l:mass-theta1-theta2} imply the result.
\hfill $\square$

 \medskip

\noindent{\bf End of the proof of  Theorem \ref{T:main_2}.} 
We have seen in the proof of Theorem \ref{T:main_1} that the mass of $\mu_{V,r}^\fR-\hat\mu_{V,r}^-$ is bounded by a constant times $r^{-d}$. Therefore, for simplicity, we will prove the first assertion in the theorem for $\hat\mu_{V,r}^-$ instead of $\mu_{V,r}^\fR$.  By \eqref{e:dist-gamma-gamma}, we can assume that $\gamma=1$. 
Let $\varphi$ be a function with compact support in $\Omega$ such that $\|\varphi\|_{\Cc^1}\leq 1$. We need to show that 
$|\langle \hat\mu_{V,r}^- -\mu_\MZ,\varphi \rangle|\lesssim r^{-\nu/2+\eta}$ for any constant $\eta>0$. 

Observe that $\Omega\cap\R$ is either empty or a finite union of open bounded intervals.
Define $\hat\Omega_-:=\Omega\cap \C_-$. We add here a hat in order to avoid any confusion. Let $\hat\Omega_+$ be the symmetry of $\hat\Omega_-$ with respect to $\R$ and let $\hat\Omega$ be the union of $\hat\Omega_+,\hat\Omega_-$ and $\Omega\cap\R$. So $\hat\Omega$ is a bounded open set, symmetric with respect to $\R$, whose boundary is piecewise smooth and transverse to $\R$.  
Define $\hat\varphi(z):=\varphi(z)$ for $z$ in $\hat\Omega_-$  or in $\Omega\cap\R$ and   $\hat\varphi(z):=\varphi(\overline z)$ if $z\in \hat\Omega_+$. Using the description of $\mu$ in Lemma \ref{l:mu}, we see that 
the desired estimate is equivalent to the inequality $|\langle \hat\mu_{V,r}^++\hat\mu_{V,r}^- -\mu,\hat\varphi \rangle|\lesssim r^{-\nu/2+\eta}$. 
 
We apply the second assertion of Lemma \ref{l:uVrhat} and Lemma \ref{l:muVrhat} for $\gamma=1$ and for large enough discs. If $\psi$ is a function with compact support in $\hat\Omega$ such that $\|\psi\|_{\Cc^1}\leq 1$, then we have
$$|\langle \hat\mu_{V,r}^++\hat\mu_{V,r}^- -\mu,\psi \rangle|\lesssim r^{-1}+ \|\hat u_{V,r}-H\|_{L^1(U)}^{1/2} \lesssim r^{-\nu/2+\eta}$$ 
for some disc $U$ large enough. 
The estimate still holds when $\psi$ is a Lipschitz function with Lipschitz constant smaller or equal to 1, $|\psi|\leq 1$,  because using the standard convolution we can approximate it uniformly by smooth functions with $\Cc^1$-norm at most equal to 1. So we can apply the identity for the above function $\hat\varphi$ instead of $\psi$ and this completes the proof of the first assertion. 
 
Observe that we can write $\Omega$ as the disjoint union of a finite number of subsets so that each of them is either a nice open set as in 
Lemma \ref{l:domain} or the union of such   a  set  with some smooth pieces of boundary.
We can also choose these sets so that these pieces of  boundaries are transverse to the real line.
Thus, for simplicity, we can assume that $\Omega$ is the nice open set described in Lemma \ref{l:domain}. Let $\hat\Omega$ be defined as above.

We apply Lemma \ref{l:dist} to $\hat\Omega$ instead of $\Omega$, and for $u_2:=\hat u_{V,r}$, $u_1:=H$, $\epsilon:=\|\hat u_{V,r}-H\|_{L^1(U)}^{1/3}$. Recall that 
$\ddc \hat u_{V,r}=\hat\mu_{V,r}^++\hat\mu_{V,r}^-+2\hat\mu_{V,r}^0$ and $\ddc H=\mu=\mu^++\mu^-+2\mu^0$. Moreover,
by Lemma \ref{l:uVrhat}, we have $\|\hat u_{V,r}-H\|_{L^1(U)}\lesssim r^{-\nu+\eta}$.
Observe also that since the boundary of $\hat\Omega$ is piecewise smooth and transverse to $\R$, we have $\mu\big((b\hat\Omega)_\epsilon\big)=O(\epsilon)$.   Therefore, applying Lemma \ref{l:dist} gives 
$$|\hat\mu_{V,r}^+(\hat\Omega)+\hat\mu_{V,r}^-(\hat\Omega)+2\hat\mu_{V,r}^0(\hat\Omega)-\mu^+(\hat\Omega)-\mu^-(\hat\Omega)-2\mu^0(\hat\Omega)|\lesssim r^{-\nu/3+\eta}$$
for every $\eta>0$, or equivalently,
$$|2\hat\mu_{V,r}^-(\Omega)+2\hat\mu^0_{V,r}(\Omega)-2\mu_\MZ(\Omega)|\lesssim r^{-\nu/3+\eta}.$$
This, together with  the mass estimate $\hat\mu_{V,r}^0(\Omega)\leq Ar^{-1}\Leb_\R(\Omega)$  obtained in Lemma \ref{l:muVrhat}, imply 
$$|\hat\mu_{V,r}^-(\Omega)-\mu_\MZ(\Omega)|\lesssim r^{-\nu/3+\eta} \quad \text{and hence}\quad 
|\mu_{V,r}^\fR(\Omega)-\mu_\MZ(\Omega)|\lesssim r^{-\nu/3+\eta}.$$
Using the last assertion in Lemma \ref{l:dist}, 
we obtain  the same estimate for $\overline\Omega$ instead of $\Omega$.

Finally, since $n_{V,W}(r)=c_da^dr^d\mu_{V,r}^\fR(W)$, the inequalities obtained above imply the last estimate in Theorem \ref{T:main_2} for $W:=\Omega$ or $\overline\Omega$. Recall that there are only finitely many resonances in $\overline\C_+$. So 
the estimate holds for all $W$ such that $\Omega\cap\C_-\subset W\subset\overline\Omega$. This completes the proof of the theorem.
\hfill $\square$

\end{document}